\documentclass[journal,comsoc]{IEEEtran}
\usepackage[T1]{fontenc}
\usepackage{cite}

\ifCLASSINFOpdf
\else
\fi
\usepackage{subfigure}
\usepackage{multicol}
\usepackage{multirow}
\usepackage{graphicx}
\usepackage{comment}
\usepackage{amsmath,amssymb} 
\usepackage{color}

\usepackage{amsmath}
\usepackage[cmintegrals]{newtxmath}
\usepackage{array}
\begin{document}
\title{RegNet: Self-Regulated Network for Image Classification}
%

\author{Jing~Xu,
        Yu~Pan,
        Xinglin~Pan,
        Steven Hoi,~\IEEEmembership{Fellow,~IEEE},
        Zhang Yi,~\IEEEmembership{Fellow,~IEEE},
        and~Zenglin~Xu$^*$.
        
\thanks{Jing~Xu and Zenglin Xu are with the School of Science and Technology, Harbin Institute of Technology, Shenzhen, Shenzhen 510085, Guangdong, China.}
\thanks{Yu~Pan and Xinglin Pan are with the Department of SMILE Lab, School of Computer Science and Engineering, University of Electronic Science and Technology of China, Chengdu 610031, China.}
\thanks{Steven~Hoi is with the School of Information Systems (SIS)
Singapore Management University, Singapore}
\thanks{Zhang~yi is with the Machine Intelligence Laboratory, College of Computer Science, Sichuan University, Chengdu 610065, China}
\thanks{Zenglin~Xu is the corresponding author (e-mail:zenglin@gmail.com)}}
\maketitle

\begin{abstract}
The ResNet and its variants have achieved remarkable successes in various computer vision tasks. Despite its success in making gradient flow through building blocks, the simple shortcut connection mechanism limits the ability of re-exploring new potentially complementary features due to the additive function. To address this issue, in this paper, we propose to introduce a regulator module as a memory mechanism to extract complementary features, which are further fed to the ResNet. In particular, the regulator module is composed of convolutional RNNs (e.g., Convolutional LSTMs or Convolutional GRUs), which are shown to be good at extracting spatio-temporal information. We named the new regulated networks as RegNet. The regulator module can be easily implemented and appended to any ResNet architectures. We also apply the regulator module for improving the Squeeze-and-Excitation ResNet to show the generalization ability of our method. Experimental results on three image classification datasets have demonstrated the promising performance of the proposed architecture compared with the standard ResNet, SE-ResNet, and other state-of-the-art architectures. 
\end{abstract}


%
\IEEEpeerreviewmaketitle

\section{Introduction}
Convolutional neural networks (CNNs) have achieved abundant breakthroughs in a number of computer vision tasks~\cite{lecun1995convolutional}. Since the champion achieved by AlexNet\cite{NIPS2012_4824} at the ImageNet competition in 2012, various new architectures have been proposed, including  VGGNet\cite{DBLP:journals/corr/SimonyanZ14a}, GoogLeNet\cite{DBLP:journals/corr/SzegedyLJSRAEVR14}, ResNet\cite{DBLP:journals/corr/HeZRS15}, DenseNet~\cite{DBLP:journals/corr/HuangLW16a}, and recent NASNet\cite{DBLP:journals/corr/ZophVSL17}. 

\begin{figure*}[tbp] 
  \centering 
  \subfigure[]{ 
    \centering 
    \includegraphics[width=0.7\textwidth,trim=0 5 0 0, clip=true]{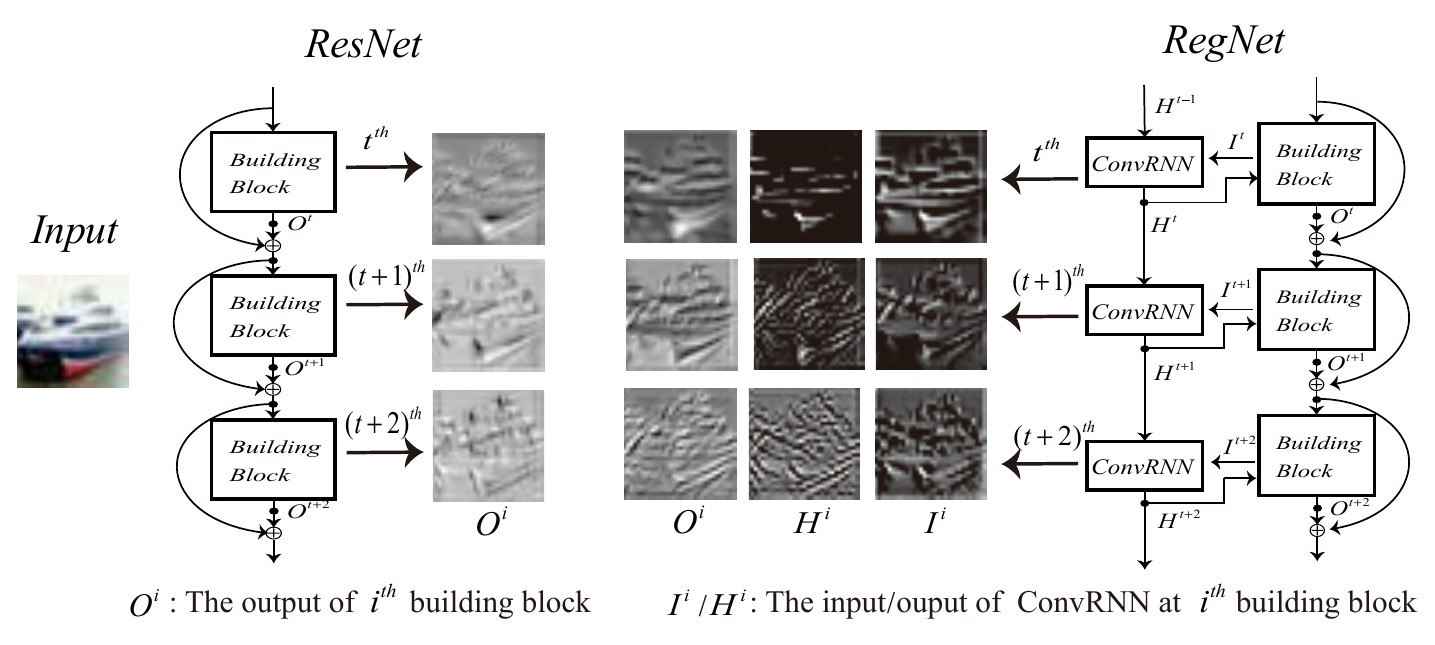} 
    \label{fig:feature_visualization}
  } 
  \subfigure[]{ 
    \centering 
    \includegraphics[width=0.25\textwidth,trim=0 0 5 5, clip=true]{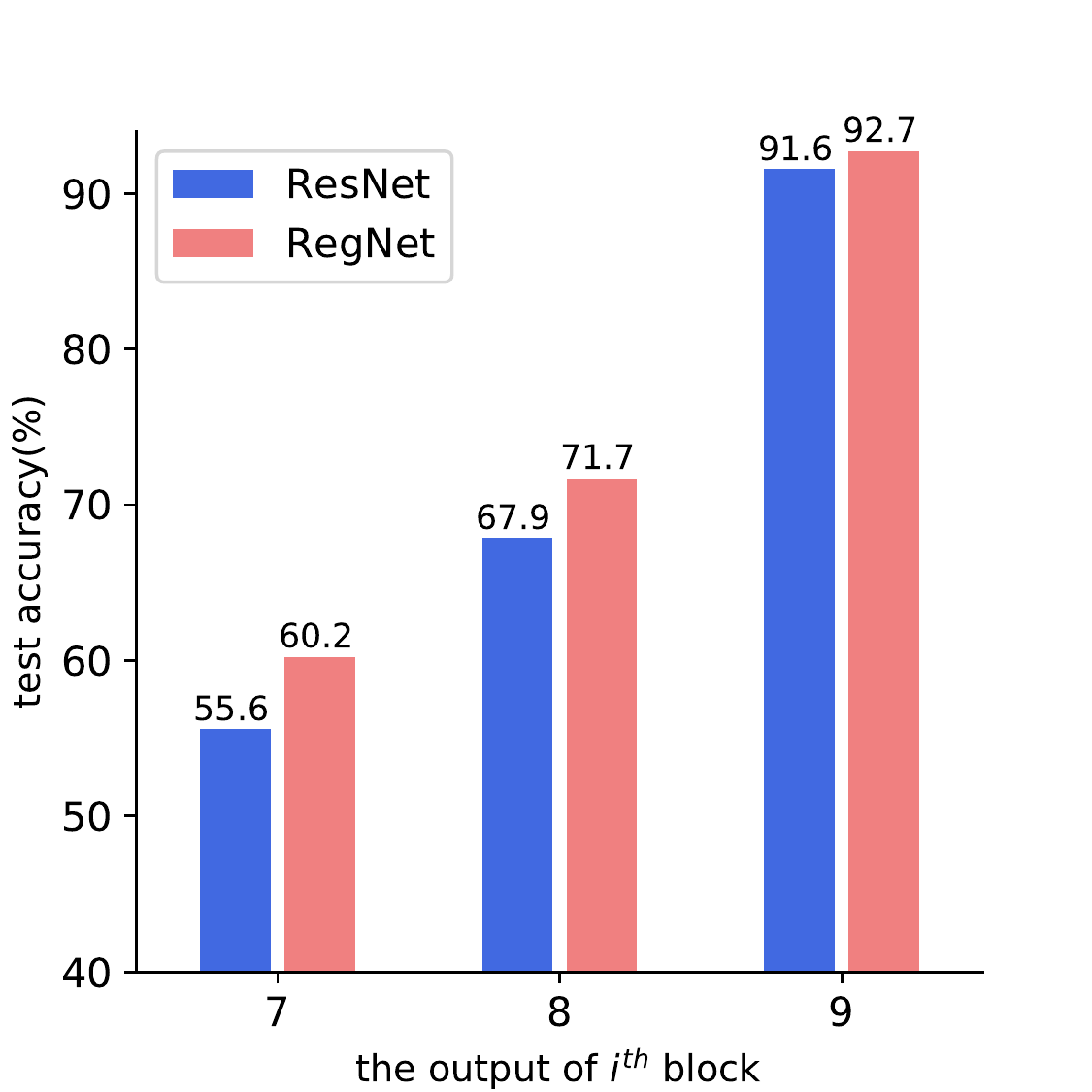} 
    \label{fig:acc_blocks}
  } 
  \caption{(a):Visualization of feature maps in the ResNet~\cite{DBLP:journals/corr/HeZRS15} and RegNet. We visualize the outputs $O^i$ feature maps of the $i^{th}$ building blocks, $i\in\{t, t+1, t+2\}$. In RegNets, $I^i$  denotes the input feature maps.  $H^i$ denotes the hidden states generated by the ConvRNN at step $i$. By applying convolution operations over the concatenation $I^i$ with $H^i$, we can get the regulated outputs( denoted by $O^i$) of the $i^{th}$ building block. (b): The prediction on test data based on the output feature maps of consecutive building blocks. During the test time, we add an average pooling layer and the last fully connected layer to the outputs of the last three building blocks($i\in\{7,8,9\}$) in ResNet-20 and RegNet-20 to get the classification results. It can be seen that the output of each block aided with the memory information results in higher classification accuracy.} 
  \label{fig:ResNet}
\end{figure*}

Among these deep architectures, ResNet and its variants~\cite{DBLP:journals/corr/ZagoruykoK16,DBLP:journals/corr/SzegedyIV16,DBLP:journals/corr/XieGDTH16,DBLP:journals/corr/abs-1709-01507} have obtained significant attention with outstanding performances in both low-level and high-level vision tasks. The remarkable success of ResNets is mainly due to the shortcut connection mechanism, which makes the training of a deeper network possible, where gradients can directly flow through building blocks and the gradient vanishing problem can be avoided in some sense. However, the shortcut connection mechanism makes each block focus on learning its respective residual output, where the inner block information communication is somehow ignored and some reusable information learned from previous blocks tends to be forgotten in later blocks. 
To illustrate this point, we visualize the output(residual) feature maps learned by consecutive blocks in ResNet in Fig.~\ref{fig:feature_visualization}. It can be see that due to the summation operation among blocks, the adjacent outputs $O^{t}$, $O^{t+1}$ and $O^{t+2}$ look very similar to each other, which indicates that less new information has been learned through  consecutive blocks.


  

A potential solution to address the above problems is to capture the spatio-temporal dependency between building blocks while constraining the speed of parameter increasing. To this end, we introduce a new regulator mechanism in parallel to the shortcuts in ResNets for controlling the necessary memory information passing to the next building block. In detail, we adopt the Convolutional RNNs (``ConvRNNs")~\cite{DBLP:journals/corr/ShiCWYWW15} as the regulator to encode the spatio-temporal memory. We name the new architecture as RNN-Regulated Residual Networks, or ``RegNet" for short. As shown in Fig.~\ref{fig:feature_visualization}, at the $i^{th}$ building block, a recurrent unit in the convolutional RNN takes the feature from the current building block as the input (denoted by $I^i$), and then encodes both the input and the serial information to generate the hidden state (denoted by $H^{i}$); the hidden state will be concatenated with the input for reuse in the next convolution operation (leading to the output feature $O^i$), and will also be transported to the next recurrent unit. To better understand the role of the regulator, we visualize the feature maps, as shown in Fig.~\ref{fig:feature_visualization}. We can see that the $H^{i}$ generated by ConvRNN can complement with the input features $I^i$. After conducting convolution on the concatenated features of $H^{i}$ and $I^i$, the proposed model gets more meaningful features with rich edge information $O^i$ than ResNet does. For quantitatively evaluating the information contained in the feature maps, we test their classification ability on test data (by adding average pooling layer and the last fully connected layer to the $O^i$ of the last three blocks). As shown in Fig.~\ref{fig:acc_blocks}, we can find that the new architecture can get higher prediction accuracy, which indicates the effectiveness of the regulator from ConvRNNs.

Thanks to the kind of parallel structure of the regulator module, the RNN-based regulator is easy to implement and can be applicable to other ResNet-based structures, such as the SE-ResNet~\cite{DBLP:journals/corr/abs-1709-01507}, Wide ResNet~\cite{DBLP:journals/corr/ZagoruykoK16},  Inception-ResNet~\cite{DBLP:journals/corr/SzegedyIV16}, ResNetXt~\cite{DBLP:journals/corr/XieGDTH16}, Dual Path Network(DPN)~\cite{DBLP:journals/corr/ChenLXJYF17}, and so on. Without loss of generality, as another instance to demonstrate the effectiveness of the proposed regulator, we also apply the ConvRNN module for improving the Squeeze-and-Excitation ResNet (shorted as ``SE-RegNet").

For evaluation, we apply our model to the task of image classification on three highly competitive benchmark datasets, including CIFAR-10, CIFAR-100, and ImageNet. 
In comparison with the ResNet and SE-ResNet, our experimental results have demonstrated that the proposed architecture can significantly improve the classification accuracy on all the datasets. We further show that the regulator can reduce the required depth of ResNets while reaching the same level of accuracy.

\section{Related Work}

Deep neural networks have been achieved empirical breakthroughs in machine learning. 
However, training networks with sufficient depths is a very tricky problem. 
Shortcut connection has been proposed to address the difficulty in optimization to some extent~\cite{DBLP:journals/corr/SrivastavaGS15,DBLP:journals/corr/HeZRS15}. Via the shortcut, information can flow across layers without attenuation. A pioneering work is the Highway Network~\cite{DBLP:journals/corr/SrivastavaGS15}, which implements the shortcut connections by using a gating mechanism. In addition, the ResNet~\cite{DBLP:journals/corr/HeZRS15} explicitly requests building blocks fitting a residual mapping, which is assumed to be easier for optimization.

Due to the powerful capabilities in dealing with vision tasks of ResNets, a number of variants have been proposed, 
including WRN~\cite{DBLP:journals/corr/ZagoruykoK16}, 
Inception-ResNet~\cite{DBLP:journals/corr/SzegedyIV16}, 
ResNetXt~\cite{DBLP:journals/corr/XieGDTH16}, 
, WResNet \cite{shen2016weighted}, 
and so on. ResNet and ResNet-based models have achieved impressive, record-breaking performance in many challenging tasks.
In object detection, 50- and 101-layered ResNets are usually used as basic feature extractors in many models: Faster R-CNN~\cite{DBLP:journals/corr/RenHG015}, RetinaNet~\cite{DBLP:journals/corr/abs-1708-02002}, Mask R-CNN~\cite{DBLP:journals/corr/HeGDG17} and so on. 
The most recent models aiming at image super-resolution tasks, such as SRResNet~\cite{DBLP:journals/corr/LedigTHCATTWS16}, EDSR and MDSR~\cite{Lim_2017_CVPR_Workshops}, are all based on ResNets, with a little modification. Meanwhile, in ~\cite{DBLP:conf/cvpr/FuHZHDP17}, the ResNet is introduced to remove rain streaks and obtains the state-of-the-art performance.

Despite the success in many applications, ResNets still suffer from the depth issue~\cite{DBLP:journals/corr/HuangSLSW16}. 
DenseNet proposed by ~\cite{DBLP:journals/corr/HuangLW16a} concatenates the input features with the output features using a densely connected path in order to encourage the network to reuse all of the feature maps of previous layers. Obviously, not all feature maps need to be reused in the future layers, and consequently the densely connected network also leads to some redundancy with extra computational costs. Recently, Dual Path Network~\cite{DBLP:journals/corr/ChenLXJYF17} and Mixed link Network~\cite{DBLP:journals/corr/abs-1802-01808} are the trade-offs between ResNets and DenseNets. In addition, some module-based architectures are proposed to improve the performance of the original ResNet. SENet\cite{DBLP:journals/corr/abs-1709-01507} proposes a lightweight module to get the channel-wise attention of intermediate feature maps. CBAM~\cite{DBLP:journals/corr/abs-1807-06521} and BAM~\cite{DBLP:journals/corr/abs-1807-06514} design modules to infer attention maps along both channel
and spatial dimensions. Despite their success, those modules try to regulate the intermediate feature maps based on the attention information learned by the intermediate feature themselves, so the full utilization of historical spatio-temporal information of previous features still remains an open problem.

On the other hand, convolutional RNNs (shorted as ConvRNN), such as 
ConvLSTM~\cite{DBLP:journals/corr/ShiCWYWW15} and ConvGRU~\cite{DBLP:journals/corr/BallasYPC15}, have been used to capture spatio-temporal information in a number of applications, such as rain removal\cite{DBLP:journals/corr/abs-1807-05698}, video super-resolution\cite{8579237}, video compression\cite{DBLP:journals/corr/abs-1910-12286}, video object detection and segemetation\cite{DBLP:journals/corr/abs-1903-10172,DBLP:journals/corr/SiamVJR16}. Most of those works embed ConvRNNs into models to capture the dependency information in a sequence of images.   
In order to regulate the information flow of ResNet, we propose to leverage ConvRNNs as a separate module aiming to extracting spatio-temporal information as complementary to the original feature maps of ResNets.

\section{Our Model}
In the section, we first revisit the background of ResNets and two advanced ConvRNNs: ConvLSTM and ConvGRU. Then we present the proposed RegNet architectures.

\subsection{ResNet}
\begin{figure}[h] 
  \centering 
  \subfigure[]{ 
  \begin{minipage}{0.15\columnwidth}
    \centering 
    \includegraphics[width=.9\textwidth,trim=0 22 0 0,clip]{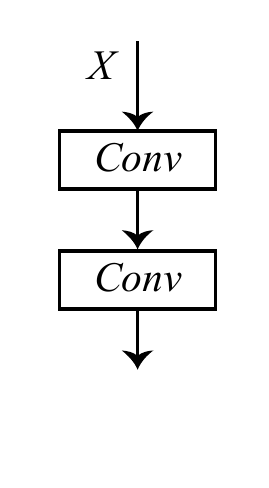} 
    \label{fig:underlying_mapping}
   \end{minipage}
  }  \quad \quad
  \subfigure[]{ 
  \begin{minipage}{0.15\columnwidth}
    \centering 
    \includegraphics[width=.9\textwidth]{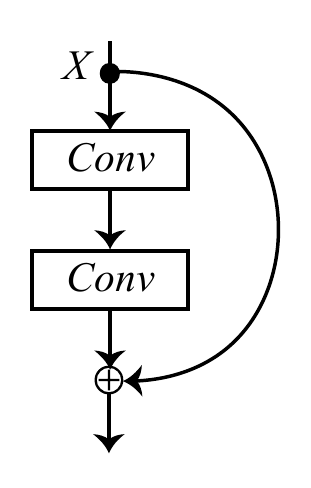} 
    \label{fig:residual_mapping}
    \end{minipage}
  }

  
  \caption{~\ref{fig:underlying_mapping} shows the original underlying mapping while~\ref{fig:residual_mapping} shows the residual mapping in ResNet~\cite{DBLP:journals/corr/HeZRS15}.
  } 
  \label{fig:ResNet}
\end{figure}

The degradation problem which makes the traditional network hard to converge, is exposed when the architecture goes deeper. The problem can be mitigated by ResNet~\cite{DBLP:journals/corr/HeZRS15} to some extent. Building blocks are the basic architecture of ResNet, as shown in Fig.~\ref{fig:residual_mapping}, instead of directly fitting a original underlying mapping, shown in Fig.~\ref{fig:underlying_mapping}. The deep residual network obtained by stacking building blocks has achieved excellent performance in image classification, which proves the competence of the residual mapping.

\subsection{ConvRNN and its Variants}

RNN and its classical variants LSTM and GRU have achieved great success in the field of sequence processing. To tackle the spatio-temporal problems, we adopt the basic ConvRNN and its variants ConvLSTM and ConvGRU, which are transformed from the vanilla RNNs by replacing their fully-connected operators with convolutional operators. Furthermore, for reducing the computational overhead, we delicately design the convolutional operation in ConvRNNs. In our implementation, the ConvRNN can be formulated as
\begin{align}
    &\mathbf{H}^t = tanh(^{2N}\mathbf W_{h}^N*[\mathbf X^t, \mathbf H^{t-1}]+\mathbf b_h),
\end{align}
where $X^t$ is the input 3D feature map, $H^{t-1}$ is the hidden state obtained from the earlier output of ConvRNN and $H^{t}$ is the output 3D feature map at this state. Both the number of input $X^t$ and output $H^{t}$ channels in the ConvRNN are N.

Additionally, $^{2N}\mathbf W^N*\mathbf X$ denotes a convolution operation between weights $\mathbf W$ and input  $\mathbf X$ with the input channel 2N and the output channel N. To make the ConvRNN more efficient, inspired by \cite{DBLP:journals/corr/abs-1903-10172,DBLP:conf/cvpr/2018},  given input $\mathbf X$ with 2N channels, we conduct the convolution operation in 2 steps:

\begin{enumerate}
\renewcommand{\labelenumi}{(\theenumi)}
\item ~Divide the input $\mathbf X$ with 2N channels into N groups, and use grouped convolutions\cite{DBLP:journals/corr/ZhangQ0W17} with $1 \times 1$ kernel to process each group separately for fusing input channels. 
\item ~Divide the feature map obtained by (1) into N groups, and use grouped convolutions with $3 \times 3$ kernel to process each group separately for capturing the spatial information per input channel. 
\end{enumerate}

Directly applying the original convolutions with $3 \times 3$ kernels suffers from high computational complexity. As detailed in Table~\ref{table:structure_compare}, the new modification reduces the required computation by 18N/11 times with comparable result. Similarly, all the convolutions in ConvGRU and ConvLSTM are replaced with the light-weight modification.

\begin{table}[t]
\caption{Performance of RegNet-20 with ConvGRU as regulators on CIFAR-10. We compare the test error rates between traditional 3$\times$3 kernels and our new modification.}
\begin{center}
\begin{tabular}{c|c|c|c}
  \hline
kernel type  & err. & Params & FLOPs \\
\hline\hline
3$\times$3   &  7.35  & +330K   & +346M   \\ \hline
              
Ours & 7.42 & +44K  & +15M    \\ \hline

\end{tabular}
\end{center}
\label{table:structure_compare}
\end{table}

\subsection{RNN-Regulated ResNet}

\begin{figure}[tbp] 
  \centering 
  \subfigure[]{
    \centering
    \includegraphics[width=.21\textwidth]{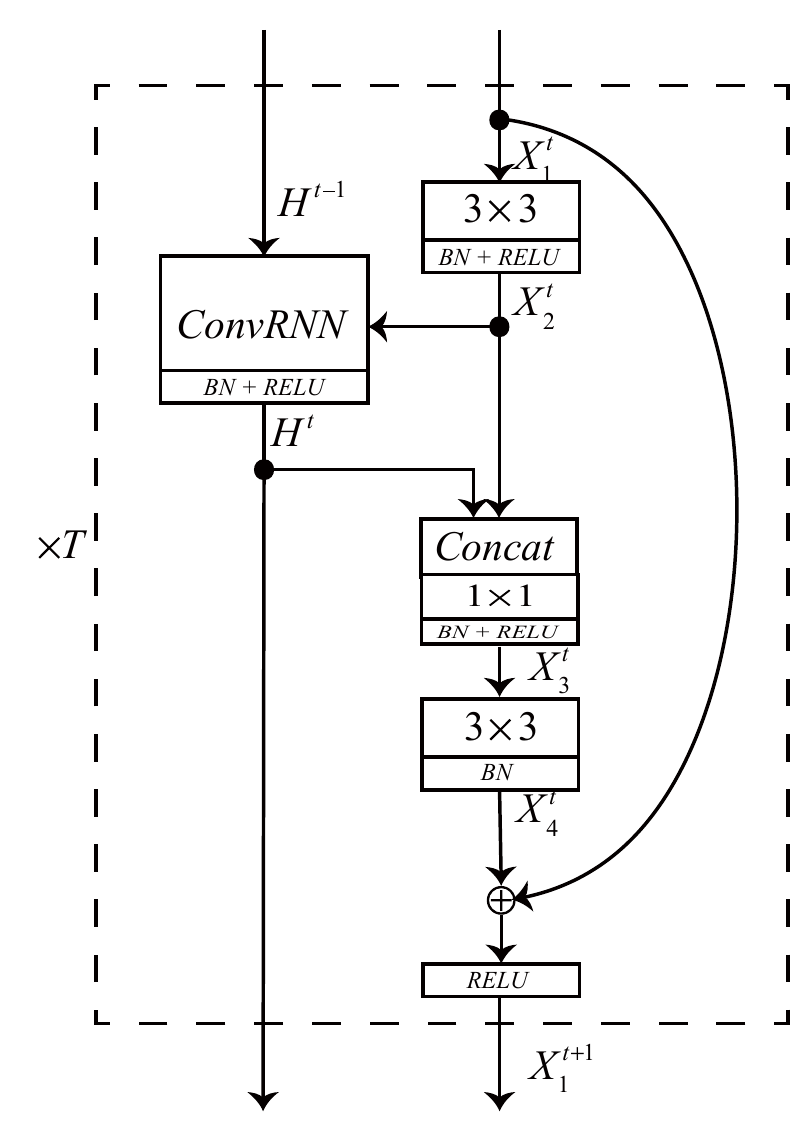} 
    \label{fig:block_a}
  } 
  \quad 
  \subfigure[]{ 
    \centering 
    \includegraphics[width=.21\textwidth]{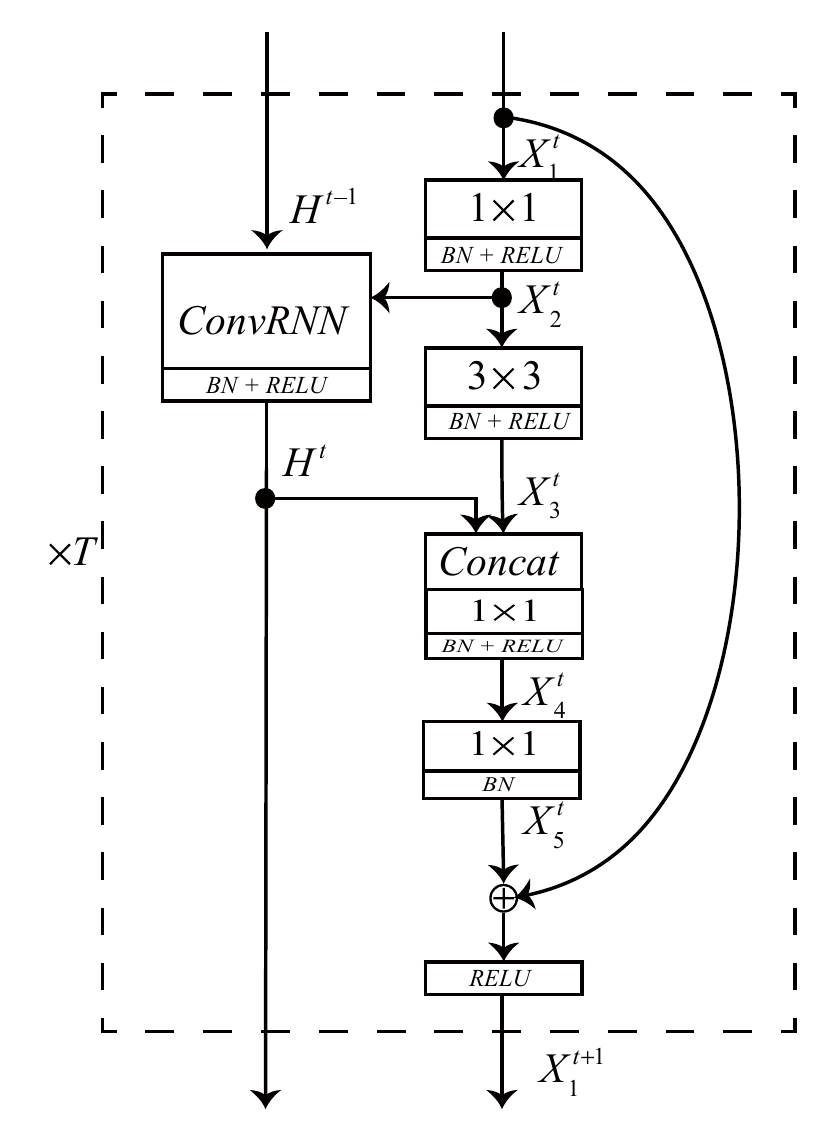} 
    \label{fig:block_c}
  } 
  \caption{The RegNet module is shown in \ref{fig:block_a}. The bottleneck RegNet block is shown in \ref{fig:block_c}. The $T$ denotes the number of building blocks as well as the total time steps of ConvRNN. }
  \label{fig:block}
\end{figure}




To deal with the CIFAR-10/100 datasets and the Imagenet dataset, \cite{DBLP:journals/corr/HeZRS15} proposed two kinds of ResNet building blocks: the non-bottleneck building block and the bottleneck building block. Based on those, by applying ConvRNNs as regulators, we get RNN-Regulated ResNet building module and bottleneck RNN-Regulated ResNet building module correspondingly.

\subsubsection{RNN-Regulated ResNet Module (RegNet module)}
The illustration of RegNet module is shown in Fig.~\ref{fig:block_a}. Here, we choose ConvLSTM for expounding. $H^{t-1}$ denotes the earlier output from ConvLSTM, and $H^t$ is output of the ConvLSTM at $t$-th module .  $X^t_i$ denotes the $i$-th feature map at the $t$-th module. 

The $t$-th RegNet(ConvLSTM) module can be expressed as
\begin{align}
    &\mathbf X^t_2 = ReLU(BN(\mathbf W^t_{12}*\mathbf X^t_1 + \mathbf{b}^t_{12}), \notag \\
    &[\mathbf H^{t}, \mathcal{C}^t ]= ReLU(BN(ConvLSTM(\mathbf X^t_2,  [\mathbf H^{t-1}, \mathcal{C}^{t-1}]))), \notag \\
    &\mathbf X^t_3 = ReLU(BN(\mathbf W^t_{23}*Concat[\mathbf X^t_2, ~\mathbf H^{t}])), \notag \\
    &\mathbf X^t_4 = BN(\mathbf W^t_{34}*\mathbf X^t_3 + \mathbf{b}^t_{34}), \notag \\
    &\mathbf X^{t+1}_1 = ReLU(\mathbf X^t_1 + \mathbf X^t_4), 
    \label{eq:o_SEResNet}
\end{align}
where $\mathbf W^t_{ij}$ denotes the convolutional kernel which mapping feature map $\mathbf X^t_i$ to $\mathbf X^t_j$ and $\mathbf b^t_{ij}$ denotes the correlative bias. Both $\mathbf W^t_{12}$ and $\mathbf W^t_{34}$ are $3\times 3$ convolutional kernels. The $\mathbf W^t_{23}$ is $1\times 1$ kernel. BN($\cdot$) indicates batch normalization. $Concat[\cdot]$ refers to the concatenate operation.

Notice that in Eq~(\ref{eq:o_SEResNet}) the input feature $\mathbf X^t_2$ and the previous output of ConvLSTM $\mathbf H^{t}$ are the inputs of ConvLSTM in $t$-th module. 
According to the inputs, the ConvLSTM automatically decides 
whether the information in memory cell will be propagated to the output hidden feature map $\mathbf H^t$.

\subsubsection{Bottleneck RNN-Regulated ResNet Module (bottleneck RegNet module)}

The bottleneck RegNet module based on the bottleneck ResNet building block is shown in Fig.~\ref{fig:block_c}. The bottleneck building block introduced in~\cite{DBLP:journals/corr/HeZRS15} for dealing with the pictures with large size.
Based on that, the $t$-th bottleneck RegNet module can be expressed as
\begin{align}
    &\mathbf X^t_2 = ReLU(BN(\mathbf W^t_{12}*\mathbf X^t_1 + \mathbf{b}^t_{12}), \notag \\
    &[\mathbf H^t, \mathcal{C}^t] = ReLU(BN(ConvLSTM(\mathbf X^t_2, [\mathbf H^{t-1}, \mathcal{C}^{t-1}]))), \notag \\
    &\mathbf X^t_3 = ReLU(BN(\mathbf W^t_{23}*\mathbf X^t_2 + \mathbf{b}^t_{23}), \notag \\
    &\mathbf X^t_4 = ReLU(BN(\mathbf W^t_{34}*Concat[\mathbf X^t_3, ~\mathbf H^t])), \notag \\
    &\mathbf X^t_5 = BN(\mathbf W^t_{45}*\mathbf X^t_4 + \mathbf{b}^t_{45}), \notag \\
    &\mathbf X^{t+1}_1 = ReLU(\mathbf X^t_1 + \mathbf X^t_5),
    \label{eq:B-RNN-R-2}
\end{align}
where $\mathbf W^t_{12}$ and $\mathbf W^t_{45}$ are the two $1\times1$ kernels, and $\mathbf W^t_{23}$ is the $3\times3$ bottleneck kernel. The $\mathbf W^t_{34}$ is a $1\times1$ kernel for fusing feature in our model.

\section{Experiments}
In this section, we evaluate the effectiveness of the proposed convRNN regulator on three benchmark datasets, including CIFAR-10, CIFAR-100, and ImageNet. We run the algorithms on pytorch. The small-scaled models for CIFAR  are trained on a single NVIDIA 1080 Ti GPU, and the large-scaled models for ImageNet are trained on 4 NVIDIA 1080 Ti GPUs.

\subsection{Experiments on CIFAR}

\begin{table}
\caption{Architectures for CIFAR-10/100 datasets. By setting n$\in\{3, 5, 7\}$, we can get the $\{20, 32, 56\}$-layered RegNet.}
\begin{center}
\begin{tabular}{c|c|c}
\hline
name     & output size & (6n$+$2)-layered RegNet \\
\hline \hline
conv\_0  & $32\times32$ & $3\times3$, 16 \\
\hline
conv\_1 & $32\times32$ &  ConvRNN$_1$ $+ \begin{bmatrix}
                3\times3, 16 \\
                3\times3, 16 
                \end{bmatrix} \times n$  \\
\hline
conv\_2 & $16\times16$ &  ConvRNN$_2$ $+ \begin{bmatrix}
                3\times3, 32 \\
                3\times3, 32 
                \end{bmatrix} \times n$  \\
\hline
conv\_3 & $8\times8$ &   ConvRNN$_3$ $+ \begin{bmatrix}
                3\times3, 64 \\
                3\times3, 64 
                \end{bmatrix} \times n$ \\
\hline
               & $1\times1$ & AP, FC, softmax  \\
\hline
\end{tabular}
\end{center}
\label{table:cifar_structual}
\end{table}

\begin{table}
\caption{Classification error rates on the CIFAR-10/100. Best results are marked in bold.}
\begin{center}
\begin{tabular}{l|c|c}
\hline
model   & C10 & C100\\
\hline\hline
ResNet-20~\cite{DBLP:journals/corr/HeZRS15} &  8.38 & 31.72  \\
RegNet-20(ConvRNN) &  7.60 & 30.03  \\
RegNet-20(ConvGRU) &  7.42 & \textbf{29.69}  \\
RegNet-20(ConvLSTM) & \textbf{7.28} & 29.81 \\
\hline

SE-ResNet-20 &  8.02 & 31.14  \\
SE-RegNet-20(ConvRNN)& 7.55  &  29.63 \\
SE-RegNet-20(ConvGRU)&  7.25 & 29.08  \\
SE-RegNet-20(ConvLSTM) & \textbf{6.98} & \textbf{29.02} \\

\hline
\end{tabular}
\end{center}
\label{table:cifar_10_classification_error}
\end{table}

\begin{table*}[t]
\caption{Test error rates on CIFAR-10/100. We use ConvGRU and ConvLSTM as regulators of ResNet. We list the increase of parameter the architectures at the right corner of the error rates. }
\begin{center}
\setlength{\tabcolsep}{1.5mm}{
\begin{tabular}{c|c|c|c|c|c|c}
\hline
 & \multicolumn{3}{c}{C-10} & \multicolumn{3}{|c}{C-100}\\ \hline
layer  & ResNet & 
+ConvGRU & +ConvLSTM& 
ResNet &
+ConvGRU& 
+ConvLSTM\\
\hline\hline
20    &   8.38   & 7.42$_{(+0.04M)}$   & 7.28$_{(+0.04M)}$  & 31.72  &  29.69$_{(+0.04M)}$  & 29.81$_{(+0.04M)}$   \\ \hline
              
32  &   7.54   & 6.60$_{(+0.06M)}$   & 6.88$_{(+0.07M)}$  & 29.86  &  27.42$_{(+0.07M)}$  & 28.11$_{(+0.07M)}$   \\ \hline

56  &   6.78   & 6.39$_{(+0.11M)}$   & 6.45$_{(+0.12M)}$  & 28.14  &  27.02$_{(+0.11M)}$  & 27.26$_{(+0.12M)}$  \\ \hline

\end{tabular}}
\end{center}
\label{table:paramter_analysis}
\end{table*}

The CIFAR datasets~\cite{Krizhevsky09} consist of RGB image with $32\times32$ pixels.
Each dataset contains 50k training images and 10k testing images. The images in CIFAR-10 and CIFAR-100 are drawn from 10 and 100 classes respectively. We train on the training dataset and evaluate on the test dataset.

 By applying ConvRNNs to ResNet and SE-ResNet, we get the RegNet, and SE-RegNet models separately. Here, we use 20-layered RegNet and SE-RegNet to prove the wide applicability of our method. The SE-RegNet building module in Fig.~\ref{fig:block_a} is used to analysis CIFAR datasets. The structural details of SE-RegNet are shown in Table~\ref{table:cifar_structual}. The inputs of the network are $32\times32$ images. In each conv$\_{i}$, ${i}\in \{1, 2, 3 \}$ layer, there are n RegNet building modules stacked sequentially, and connected together by a ConvRNN. In summary, there are 3 ConvRNNs in our architecture, and each ConvRNN impacts on the n RegNet building modules.  The reduction ratio r in SE block is 8. 

In this experiment, we use SGD with a momentum of 0.9 and a weight decay of 1e-4. We train with a batch size of 64 for 150 epoch. The initial learning rate is 0.1 and divided by 10 at 80 epochs. Data augmentation in~\cite{DBLP:conf/aistats/2015} is used in training. The results of SE-ResNet on CIFAR are based on our implementation, since the results were not reported in~\cite{DBLP:journals/corr/abs-1709-01507}.

\subsubsection{Results on CIFAR}

The classification errors on the CIFAR-10/100 test sets are shown in Table~\ref{table:cifar_10_classification_error}.  We can see from the results, with the same layer, both RegNet and SE-RegNet outperform the original models by a significant margin. Compared with ResNet-20, our RegNet-20 with ConvLSTM decreases the error rate by 1.51\% on CIFAR-10 and 2.04\% on CIFAR-100. At the same time, compared with SE-ResNet-20, our SE-RegNet-20 with ConvLSTM decreases the error rate by 1.04\% on CIFAR-10 and 2.12\% on CIFAR-100. Using ConvGRU as the regulator can reach the same level of accuracy as ConvLSTM. Due to the vanilla ConvRNN lacks gating mechanism, it performs slightly worse but still makes great progress compared with the baseline model.

\subsubsection{Parameters Analysis}
\begin{figure}[t] 
  \centering 
  \subfigure[]{ 
    \centering 
    \includegraphics[width=.215\textwidth]{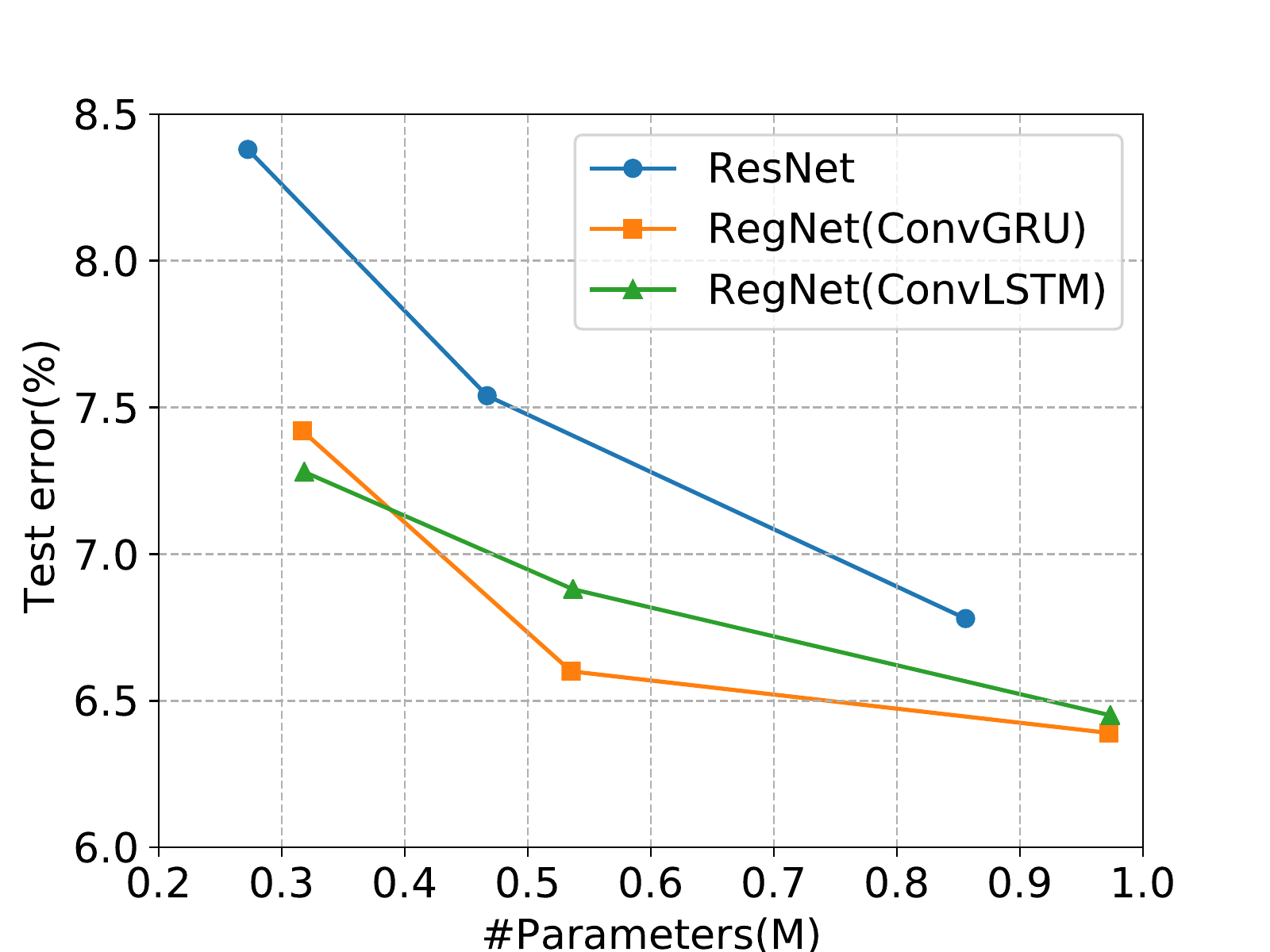} 
    \label{fig:c10}
  } 
  \subfigure[]{ 
    \centering 
    \includegraphics[width=.215\textwidth]{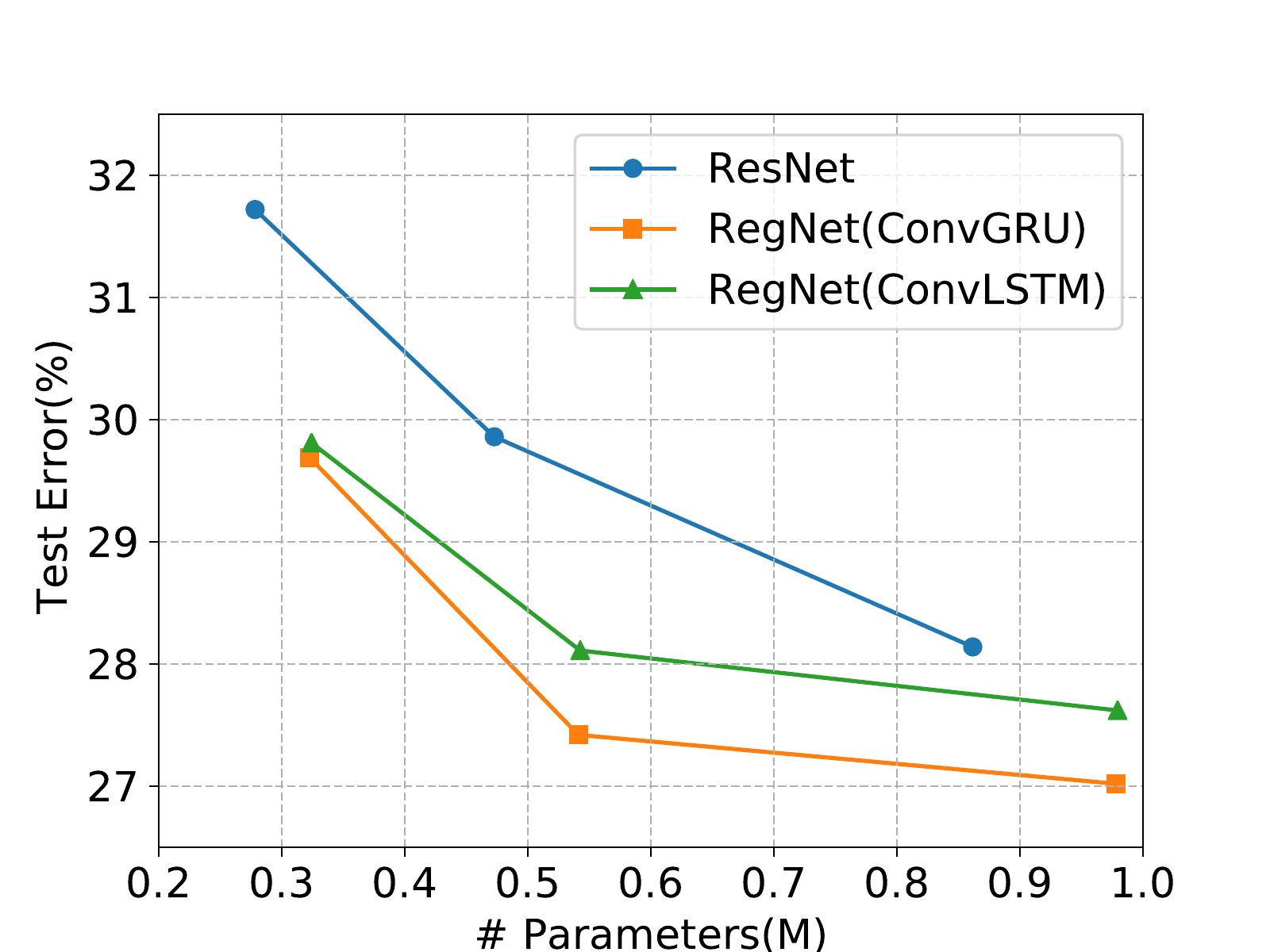} 
    \label{fig:c100}
  } 
  \caption{Comparison of parameter efficiency on CIFAR-10 between RegNet and ResNet~\cite{DBLP:journals/corr/HeZRS15}. In both \ref{fig:c10} and \ref{fig:c100}, the curves of our RegNet is always below ResNet~\cite{DBLP:journals/corr/HeZRS15} which show that with the same parameters, our models have stronger ability of expression.
  } 
  \label{fig:parameters_analysis}
\end{figure}

For a fair comparison, we evaluate our model's ability by regarding the number of models parameters as the contrast reference. As shown in Table~\ref{table:paramter_analysis}, we list the test accuracy of 20, 32, 56-layered ResNets and their respective RegNet counterparts on CIFAR-10/100. After adding minimal additional parameters, both our RegNet with ConvGRU and ConvLSTM surpass the ResNet by a 
large margin. Our 20-layered RegNet with extra 0.04M parameters even outperforms the 32-layered ResNet on both CIFAR-10/100: our 20-layered RegNet(ConvLSTM) having 0.32M parameters reaches 7.28\% error rate on CIFAR-10 surpass the 32-layered ResNet with 7.54\% error rate which having 0.47M parameters. Fig.~\ref{fig:parameters_analysis} demonstrates the parameter efficiency comparisons between RegNet and ResNet. We show our RegNet are more parameter-efficient than simply stacking layers in vanilla ResNet. On both CIFAR-10/100, our RegNets(GRU) get comparable performance with ResNet-56 with nearly 1/2 parameters.

\subsubsection{Positions of Feature Reuse}
 In this subsection, we perform ablation experiment to further analyze the effect of the position of feature reuse. We conduct an experiment to analysis that with ConvRNN which layer has the maximum promotion to the final outcome. Some previous studies~\cite{DBLP:journals/corr/YosinskiCBL14} show that the features in an earlier layer are more general while the features in later layers exhibit more specific. As shown in Table~\ref{table:cifar_structual}, the conv\_1, conv\_2, conv\_3 layers are separated by the down sampling operation, which makes the features in conv\_1 are more low-level and in conv\_3 are more specific for classification.  The classification results are shown in Table~\ref{table:rnn_pos}. In each model, only one ConvRNN is applied. We name the models RegNet$_{(i)}$, ${i}\in\{1, 2, 3\}$ which denotes that only applying a ConvRNN in layer conv$\_i$ and maintaining the original ResNet structure in the other layers. For a fair comparison, we evaluate the models ability by regarding the number of models parameters as the contrast reference. We can see from the results, using ConvRNNs in a lower layer(conv\_1) is more parameter-efficient than higher layer(conv\_3). With less parameter increasing in lower layers, they can bring about nearly same improvement in accuracy compared with higher layers. Compared with ResNet, our RegNet$_{(1)}$(GRU) decrease the test error from 8.38\% to 7.52\%(-0.86\%) on CIFAR-10 with additional 0.006M parameters and from 31.72\% to 30.40\%(-1.32\%) on CIFAR-100 with additional 0.007M parameters. This significant improvement with minimal additional parameters further proves the effectiveness of the proposed method. The concatenate operation in our model can fuse features together to explore new features~\cite{DBLP:journals/corr/ChenLXJYF17}, which is more important for general features in lower layers. 

\begin{table}[t]
\caption{Test error rates on CIFAR-10/100. We use ConvGRU and ConvLSTM as regulators of ResNet. We list the increase of parameter the architectures. In each of our RegNet$_{(i)}$ models, there is only one ConvRNN applied in layer conv$\_i$, ${i}\in\{1, 2, 3\}$. }
\begin{center}
\begin{tabular}{l|c|c|c|c}
\hline
 & \multicolumn{2}{c}{C-10} & \multicolumn{2}{|c}{C-100}\\ \hline
model  &~~ err.~~ &~~ Params~~ &~~ err. ~~& ~~Params~~\\
\hline\hline
ResNet~\cite{DBLP:journals/corr/HeZRS15}   &   8.38   & 0.273M   & 31.72  & 0.278M  \\ \hline
              
RegNet$_{(1)}$(GRU) & 7.52 & 0.279M  & 30.40 & 0.285M     \\ \hline

RegNet$_{(2)}$(GRU) & 7.48 &   0.285M      & 30.34  &  0.291M    \\ \hline

RegNet$_{(3)}$(GRU)  & 7.49 &   0.306M     & 30.30  &  0.312M  \\ \hline

RegNet$_{(1)}$(LSTM)  & 7.56 & 0.281M & 30.23 &  0.286M    \\ \hline

RegNet$_{(2)}$(LSTM)  &   7.49   & 0.290M  & 30.28 & 0.296M    \\ \hline

RegNet$_{(3)}$(LSTM)  &   7.52   & 0.325M  & 29.92  & 0.331M    \\ \hline

\end{tabular}
\end{center}
\label{table:rnn_pos}
\end{table}




\begin{table}[t]
\caption{Single-crop validation error rates on ImageNet and complexity comparisons. Both ResNet and RegNet are 50-layer. ResNet$^*$ means we reproduce the result by ourself. }
\begin{center}
\begin{tabular}{l|c|c|c|c}
\hline

model &  top-1 err. & top-5 err. &Params &FLOPs\\
\hline\hline
ResNet~\cite{DBLP:journals/corr/HeZRS15}  & 24.7 & 7.8 & \multirow{2}{*}{26.6M} & \multirow{2}{*}{4.14G}\\
\cline{1-3}
ResNet$^*$& 24.81 & 7.78 &  &       \\ \hline
RegNet & 23.43$_{(-1.38)}$ &6.93$_{(-0.85)}$&31.3M &5.12G     \\ \hline

\end{tabular}
\end{center}
\label{table:imagenet_para}
\end{table}

\begin{table}[h]
\caption{Single-crop error rates on the ImageNet validation set for state-of-the-art models. The ResNet-50$^*$ means that the re-implemention result by our experiments. 
}
\begin{center}
\begin{tabular}{l|c|c|c|c}
\hline
model     & top-1 & top-5 &Params(M) &FLOPs(G)\\
\hline \hline

WRN-18(widen=2.0)~\cite{DBLP:journals/corr/ZagoruykoK16}&25.58   & 8.06 &45.6 & 6.70\\

\hline

DenseNet-169~\cite{DBLP:journals/corr/HuangLW16a}  &  23.80 & 6.85 &28.9&7.7 \\
\hline
SE-ResNet-50~\cite{DBLP:journals/corr/abs-1709-01507}  & 23.29 & 6.62 & 26.7 & 4.14\\
\hline
\hline

ResNet-50~\cite{DBLP:journals/corr/HeZRS15}  & 24.7    &  7.8 &- & -\\
ResNet-50$^*$  & 24.81    &  7.78 &26.6& 4.14  \\
ResNet-101~\cite{DBLP:journals/corr/HeZRS15}  & 23.6    & 7.1 & 44.5 &7.51 \\
RegNet-50 &\textbf{23.43}  &\textbf{6.93} &31.3&5.12  \\

\hline
\end{tabular}
\end{center}
\label{table:imagenet_classification}
\end{table}

\subsection{Experiments on ImageNet}




We evaluate our model on ImageNet 2012 dataset~\cite{DBLP:journals/corr/SimonyanZ14a} which consists of 1.28 million training images and 50k validation images from 1000 classes. Following the previous papers, We report top-1 and top-5 classification errors on the validation dataset.
Due to the limited resources of our GPUs and without of loss of generality, we  run the experiments of ResNets and RegNets only.

The bottleneck RegNet building modules are applied to ImageNet. We use 4 ConvRNNs in RegNet-50. The ConvRNN$_{i}$, ${i}\in\{1, 2, 3, 4\}$, controls $\{3, 4, 6, 3\}$ bottleneck RegNet modules respectively. In this experiment, we use SGD with a momentum of 0.9 and a weight decay of 1e-4. We train with batch size 128 for 90 epoch. The initial learning rate is 0.06 and divided by 10 at 50 and 70 epochs. The input of the network is $224\times224$ images, which randomly cropped from the resized original images or their horizontal flips. Data augmentation in ~\cite{DBLP:journals/corr/abs-1807-05698} is used in training. We evaluate our model by applying a center-crop with $224\times224$.

We evaluate the efficiency of baseline models ResNet-50 and its respectively RegNet counterpart. The comparison is based on the computational overhead. As shown in Table~\ref{table:imagenet_para} with additional 4.7M parameters, RegNet outperforms the baseline model by 1.38\% on top-1 accuracy and 0.85\% on top-5 accuracy.


Table~\ref{table:imagenet_classification} shows the error rates of some state-of-the-art models on the ImageNet validation set. Compared with the baseline ResNet, our RegNet-50 with 31.3M parameters and 5.12G FLOPs not only surpasses the ResNet-50 but also outperforms ResNet-101 with 44.6M parameters and 7.9G FLOPs.
Since the proposed regulator module is essentially a beneficial makeup to the short cut mechanism in ResNets, one can easily apply the  regulator module to other ResNet-based models, such as SE-ResNet, WRN-18~\cite{DBLP:journals/corr/ZagoruykoK16}, ResNetXt~\cite{DBLP:journals/corr/XieGDTH16}, Dual Path Network (DPN)~\cite{DBLP:journals/corr/ChenLXJYF17}, etc.
Due to computation resource limitation, we leave the implementation of the regulator module in these ResNet extensions as our future work. 

\section{Conclusions}

In this paper, we proposed to employ a regulator module with Convolutional RNNs to extract complementary features for improving the representation power of the ResNets. 
Experimental results on three image-classification datasets have demonstrated the promising performance of the proposed architecture in comparison with standard ResNets and Squeeze-and-Excitation ResNets as well as other state-of-the-art architectures.

In the future, we intend to further improve the efficiency of the proposed architecture and to apply the regulator module to other ResNet-based architectures \cite{DBLP:journals/corr/ZagoruykoK16,DBLP:journals/corr/SzegedyIV16,DBLP:journals/corr/XieGDTH16} to increase their capacity. Besides, we will further explore RegNets for other challenging tasks, such as object detection\cite{DBLP:journals/corr/RenHG015,DBLP:journals/corr/abs-1708-02002}, image super-resolution\cite{DBLP:journals/corr/LedigTHCATTWS16,Lim_2017_CVPR_Workshops}, and so on.


%


\section*{Acknowledgment}

This work was partially supported by the National Key Research and Development Program of China (No. 2018AAA0100204).

\ifCLASSOPTIONcaptionsoff
  \newpage
\fi



\bibliographystyle{IEEEtran}
\bibliography{IEEEabrv}
\end{document}